\documentstyle[aps,twocolumn,psfig]{revtex}
\begin{document}
\author{N.K. Kuzmenko\\V.G. Khlopin Radium Institute, 194021,
St.-Petersburg, Russia\\
V.M. Mikhajlov\\
Institute of Physics St.--Petersburg State
University 198904, Russia\\
and S. Frauendorf \\
Department of Physics, University of Notre Dame,
Notre Dame, IN 46556, USA and  IKH, Research Center
Rossendorf, Germany 
}
\title{Step wise destruction of the pair correlations in micro clusters
by a magnetic field}
\maketitle

\begin{abstract}
The response of $nm$-size spherical superconducting clusters
to a magnetic field is  studied for the canonical ensemble of
electrons in a single degenerate shell. For temperatures close to zero,
the discreteness of the electronic
states causes a step like destruction of the pair correlations
 with increasing field strength, which shows up as
 peaks in the susceptibility and heat capacity.
 At higher temperatures the transition  becomes
smoothed out and extends to field strengths where the pair correlations
are destroyed at zero temperature.
\end{abstract}

The electron pair correlations in small systems
where the single-particle spectrum is discrete and the mean
level spacing is comparable with  the pairing gap have recently
been studied by means  of electron transport through
$nm$-scale Al clusters \cite{3}. The pair correlations were found to
sustain an external magnetic field of several Tesla, in contrast to
much weaker critical field of $H_c=99$ Gauss of   bulk  Al. A step wise
destruction of the pair correlations was suggested \cite{3}. It is
caused by  the subsequent excitation of quasi particle levels, which gain
energy due to the interaction of their spin with  the external field.
This mechanism is very different from the transition to the normal state
caused by a magnetic field applied to a macroscopic superconductor.
Hence it is
expected that  physical quantities, as the susceptibility $\chi$ and the
specific heat capacity $C$, which indicate the transition, behave very
differently in the micro cluster. The present letter addresses this
question.

For the micro clusters the energy to remove an electron is much larger than 
the temperature $T$. The fixed number of the electrons on the cluster was
demonstrated by the tunneling experiments \cite{3}.
Hence, one must study the transition from the paired to the unpaired state
in the frame of the canonical ensemble.  The small number
of particles taking part in superconductivity causes
 considerable fluctuations of the order parameter, which modify the
transition \cite{muehl,kuzmenko}. Consequences of particle number
conservation for the pair correlations in micro clusters have also been
discussed recently in \cite{5,6,golubev,mastellone,braun}, where a more
complete list of references to earlier work can be found.

In order to elucidate the qualitative features we consider the highly
idealized model of pair correlations between electrons in a degenerate
level, which permits to calculate the canonical partition function.
A perfect Al-sphere of radius  $R=(1-5)nm$ confines
$N\approx 2\cdot 10^3- 3\cdot10^5$ free electrons. 
Its electron levels  have good angular
momentum $l$. Taking the electron spin into account, each of these levels
has a degeneracy of $2M$, where $M=2l+1$. 
For a spherical oscillator potential, the average angular momentum at the 
Fermi surface is $\l\approx N^{1/3}-1.4$, where  
  $N$ is the number 
of  free electrons. The distance
between these levels is $\Delta e \sim 10$ $meV$ is much larger than
the BCS gap parameter $\Delta$, which is less than 1$meV$. Therefore, 
 it is sufficient to
consider the pair correlations within the last incompletely filled level.
This single-shell model also  applies to a hemisphere, because its
spectrum consists of the spherical levels with odd $l$, and
to clusters with a
superconducting layer covering an insulating sphere (cf. \cite{layer})
or hemisphere.

The single-shell model Hamiltonian
\begin{eqnarray}\label{ham}
H=H_{pair}-\omega (L_z+2S_z), \\
H_{pair}=-GA^+A,\; \;
A^+=\sum_{k>0} a^+_ka^+_{\bar {k}}, \nonumber
\end{eqnarray}
consists of the pairing interaction $H_{pair}$, which
acts between the electrons in the last shell
with the effective strength $G$,
and the interaction with the  magnetic field. We introduced the
Larmour frequency  $\hbar\omega=\mu_B B$, the Bohr magneton $\mu_B$,
  the $z$-components of the total
orbital angular momentum and spin $L_z$ and $S_z$.
The label $k=\{\lambda ,\sigma \}$  denotes the
$z$ -projections of orbital  momentum and spin of the electrons,
respectively, and
 $A^+$   creates
an electron pair on  states
$(k,\bar {k})$, related by the  time reversal.

 The magnetic  susceptibility and
heat capacity of the electrons are
\begin{equation}\label{chi}
\chi=-\frac{\mu_B^2}{\hbar^2V}\frac{\partial^2F(T,\omega)}{\partial\omega^2},
~~C=-\frac{\partial^2F(T,\omega)}{\partial T^2},
\end{equation}
The free energy $F$ derived from the Hamiltonian
(\ref{ham}) gives only the
paramagnetic part $\chi_P$ of the susceptibility, 
because we left out the term quadratic
in $B$.  For the fields we are
interested in (magnetic length is small as compared to the cluster size),
the latter  can be treated in first order perturbation theory,
 generating the diamagnetic part of the susceptibility
\begin{equation}\label{chidia}
\chi_D=-\frac{m\mu_B^2<x^2+y^2>}{\hbar^2 V}\sim-4\cdot 10^{-6}N^{2/3},
\end{equation}
where  $m$ is  the electron mass is and  $V$ the volume of the cluster.
It is nearly temperature and field independent \cite{kuzmenko}.
The numerical estimate for Al assumes constant electron density. 
For $nm$ scale clusters  
$\chi_D\sim 10^{-3}$. It is much smaller than  $\chi_D\sim -1$ for 
 macroscopic superconductors, which show the Meissner effect.  
Since the magnetic field penetrates
the cluster it can sustain a very high field of $B\sim Tesla$.
 On the other hand, the $\chi_D$ is three orders of magnitude
larger than the Landau diamagnetic susceptibility observed in normal bulk
metals.  

The exact solutions to the pairing problem of particles in a degenerate shell
were found in nuclear physics \cite{9}
in terms of representations of the group $SU_2$.
The eigenvalues $E_{\nu}$ of $H_{pair}$ are
\begin{equation}\label{Enu}
E_{\nu}=-\frac{G}{4}(N_{sh}-\nu)(2M+2-N_{sh}-\nu).
\end{equation}
where  $N_{sh}$ is the number of  particles in the shell.
The seniority, which  is the number of
unpaired particles, is constrained by $0 \le \nu \le N_{sh}$ and $\nu\leq M$.
The degenerate states $\{\nu,i\}$ of given seniority $\nu$
 differ by their magnetic moments $\mu_Bm_{\nu,i}$, where $i=\{L\Lambda
S\Sigma\}$ takes  all values
of the total orbital $(L)$ and total spin $(S)$ momenta and their total
$z$-projections $(\Lambda ,\Sigma)$ that
are compatible with the Pauli principle
for   $\nu$ electrons.
In presence of a magnetic field
 the states  have the   energy
\begin{equation}\label{Unu}
U_{\nu,i}(\omega)=E_{\nu} - \omega m_{\nu,i},
~~m_{\nu,i}=(\Lambda+2\Sigma)_{\nu,i},
\end{equation}                                      
and the canonical partition function 
 becomes
\begin{eqnarray}\label{Z}
Z=\sum_{\nu,i}\exp(-\beta U_{\nu,i})\nonumber\\
=\sum_{\nu} \exp(-\beta E_{\nu})
[\Phi_{\nu}-\Phi_{\nu-2}(1-\delta_{\nu .0})], \\
\beta=1/T,~~~~\Phi_{\nu}=\sum_{i}
\exp(-\beta \omega m_{\nu ,i}).\label{Znu}
\end{eqnarray}
To evaluate the sums
we take into account symmetry of
the wave functions of the $\nu$ unpaired electrons  and reduce the sums
 (\ref{Znu})
 to  products of sums over orbital projections of completely
antisymmetric states (one column Young diagram, cf. \cite{Hammermesh})
with $\nu /2+\Sigma$ and $\nu /2 - \Sigma$ electrons.
\begin{eqnarray}\label{Phi}
\Phi_{\nu}=\sum_{\Sigma=\Sigma_{min}}^{\nu /2}
2(1+\delta_{\Sigma .0})^{-1}
\tilde\Phi_{\nu/2+\Sigma}\tilde\Phi_{\nu/2-\sigma}
\cosh(2\beta\omega\Sigma), \\
\Sigma_{min}=[1-(-)^{\nu}]/4 \nonumber\\
\tilde\Phi_k=\delta_{k.0}+(1-\delta_{k.0})
\prod_{\mu=1}^k\frac{\sinh\beta\omega
\frac{2l+2-\mu}{2}}{\sinh\beta\omega \frac{\mu}{2}}.
\label{Phit}
\end{eqnarray}
The derivation of (\ref{Z} - \ref{Phit}) will be published separately
\cite{oddeven}.

The pair correlation energy is
\begin{eqnarray}\label{epair}
E_c(T,\omega)=\frac{1}{Z}\sum_{\nu}E_{\nu}exp(-\beta E_{\nu})
[\Phi_{\nu}-\Phi_{\nu-2}(1-\delta_{\nu .0})]  \nonumber\\
\equiv-\Delta_c^2(T,\omega)/G.
\end{eqnarray}
Here we have defined  the parameter $\Delta_c$, which measures the amount
of  pair correlations.
Applying the mean field approximation and the grand canonical ensemble
to our model, the thus introduced $\Delta_c$   becomes the familiar BCS gap
parameter $\Delta$. Accordingly
we also refer to $\Delta_c$ as the "canonical gap". However, $\Delta_c$
 must be clearly distinguished from $\Delta$ because it incorporates the
correlations caused by the fluctuations of the order parameter $\Delta$.
For the case of a half filled shell and even $N_{sh}$,
the BCS gap is $\Delta(0)\equiv\Delta(T=0,\omega=0)=GM/2$. 
 Ref. \cite{3} found $\Delta(0)=0.3-0.4~meV$
for Al-clusters  with $R=5-10~nm$,
which sets the energy scale.

 \begin{figure}
\noindent
\mbox{\psfig{file=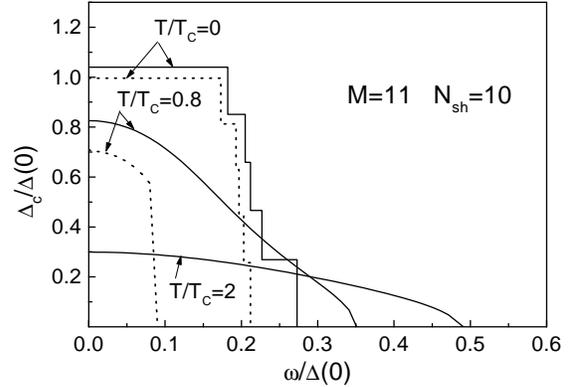,width=8cm}}
\caption{\label{f:delom}
 Canonical gap  $\Delta_c(T,\omega)$ (full lines) and BCS gap
$\Delta(T,\omega)$ (
dotted lines) v.s. the Larmour frequency $\omega$.
}
\end{figure}

Let us first consider the destruction of the pair correlations at $T=0$.
 The lowest state of each seniority multiplet
has the maximal magnetic moment
\begin{equation}\label{mnu}
m_{\nu}=\frac{\mu_B}{4} \{ \nu(2M-\nu)-
\frac{1}{2}[1-(-)^{\nu}] + 4(1-\delta_{\nu.0} ) \}.
\end{equation}
 According to (\ref{Enu}), (\ref{Unu})
and (\ref{mnu}) the state of
lowest energy changes from $\nu$ to $\nu +2$ at
\begin{eqnarray}
\omega_{\nu+2}=\frac{2\Delta (0)}{M}\left[\delta_{\nu.0} +
\frac{M-\nu}{M-\nu-1}(1-\delta_{\nu.0})\right].
\end{eqnarray}                                      
At each such step $m_{\nu}$ increases according to (\ref{mnu}).
The pair correlations
 are reduced because  two electron states are blocked.
At the last step  leading to the maximum seniority $\nu_{max}$
 all electron states are blocked. Hence the
field  $B_{c}$ corresponding to $\omega_{c}=\omega_{\nu_{max}}$ can be
regarded as
the critical one, which  destroys the pairing completely at $T=0$.
For a half filled shell 
$ \omega_{c}=3\Delta(0)/M$
for  even electron number and $4\Delta(0)/M$  for odd 
($\nu_{max}=M-1$ and $M$, respectively).

Fig. \ref{f:delom} illustrates the step wise destruction of pairing by
blocking for the half filled shell $M=11$ ($l=5$).
This mechanism was discussed in
\cite{3}, where the crossing of states  with different seniority could be
observed. It is  a well established effect in nuclear physics, where the
states of maximum angular momentum, are observed as "High-K isomers"
\cite{highk}.  Fig. \ref{f:delom} also shows   results for the mean field
(BCS) approximation (cf. \cite{kuzmenko,9}) to the single shell model.
The pair correlations are more rapidly destroyed. The
quantum fluctuations of the order parameter stabilize the pairing.

\begin{figure}[t]
\noindent
\mbox{\psfig{file=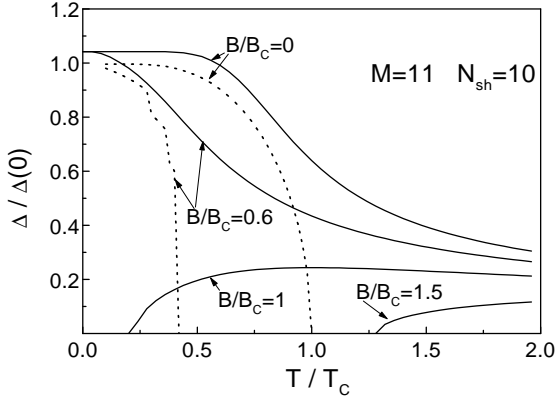,width=8cm}}
\caption{\label{f:delt}
 Canonical gap  $\Delta_c(T,\omega)$ (full lines) and BCS gap
$\Delta(T,\omega)$ (
dotted lines) v.s.
 the temperature $T$}
\end{figure}

We introduce $T_c$
as the temperature at which  the  {\em  mean field
 pair gap} $\Delta(T_c,\omega=0)$
  takes the value zero when the  magnetic field is absent.
For the half filled shell $T_c=\Delta(0)/2$.
Fig. \ref{f:delom} shows
the mean field  gap $\Delta(T,\omega)$, It
 behaves as expected from macroscopic
superconductors: The frequency where $\Delta=0$
shifts towards smaller values   with increasing $T$.
Fig. \ref{f:delt} shows that the temperature where
 $\Delta=0$ shifts from $T_c$ to lower values for  $\omega>0$.

However, fig. \ref{f:delom} also demonstrates that the canonical gap
$\Delta_c$ behaves differently.  For
$T=0.8T_c$ there is a region above $\omega_{c}$ where there are still
pair correlations. For $T=2T_c$ this region
extends to $2\omega_{c}$. The pair correlations fall off
very gradually with $\omega$. Fig. \ref{f:delt} shows how these
"temperature induced" pair correlations manifest themselves with increasing
$T$. For $\omega=0$ there is a pronounced drop of of $\Delta_c$ around
$T_c$, which signalizes the break down of the static pair field. Above this
temperature there is a long tail of dynamic pairing.
For $\omega\geq \omega_c$ the dynamic pair correlations only built up with
increasing $T$.

The temperature induced pairing can be understood in the following way:
At $T=0$,  all electrons are unpaired    when the
state of maximum seniority becomes the ground state for
$\omega>\omega_{crit}$. At $T>0$ excited states with lower seniority enter
the canonical ensemble, reintroducing the pair correlations.

 Here we have adopted the
terminology of nuclear physics, calling "static" the  mean field
(BCS) part of the pair correlations and "dynamic" the quantal and
statistical
fluctuations of the mean field  (or equivalently of the order parameter).
The "pair vibrations", which are oscillations of
the pair field around
zero  \cite{bm2}, are well established in nuclei.
 Fluctuation induced superconductivity was
discussed before  \cite{fluctuation}.
The fluctuations play a particularly important role 
in the  systems  the size of which is smaller than the
coherence length,

 \begin{figure}[t]
\noindent
\mbox{\psfig{file=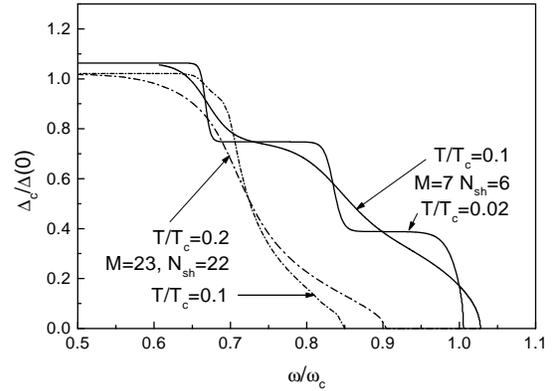,width=8cm}}
\caption{\label{f:delom723}
Canonical gap $\Delta_c (T,\omega)$ v.s.  $\omega$. }
\end{figure}

\begin{figure}[t]
\noindent
\hspace{.5cm}	
\mbox{\psfig{file=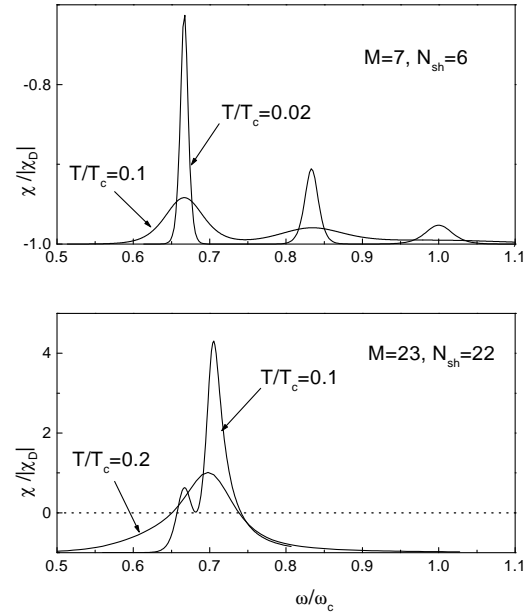,width=7cm}}
\caption{\label{f:chiom723} Susceptibility $\chi(T,\omega)$ v.s.   $\omega$. }
\end{figure}

Fig. \ref{f:delom723} shows  a very small cluster ($M=7$) with very
pronounced steps. Already at $T=0.1T_c$ the steps are noticeably washed out.
In the single-shell model the step length is $\omega_{\nu}-\omega_{\nu-2}\sim
\Delta(0)/M$. Accordingly,
no individual steps are recognizable at $T=0.1T_c$
for the large cluster  ($M=23$) shown.  Yet there is
some irregularity  around $\omega=0.7\omega_{c}$, which is a residue of
the discreteness of the electronic states. It is
thermally averaged out for $T=0.2T_c$.
Hence only for   $M<50$ i. e.  $N<2\cdot10^4$ and $T<0.1T_c$
the step wise  change  of $\Delta_c$ is observable.

The discreteness of the electronic levels has dramatic consequences for the
the susceptibility at low temperatures. As shown in the upper panel of
fig. \ref{f:chiom723}, $\chi_P$ has pronounced peaks at the frequencies where
the states with higher seniority and magnetic  moment take over. The
paramagnetic contribution is very sensitive to the temperature and
 to  the fluctuations of the order parameter.
Using the BCS mean field approximation we find much more narrow peaks,
which are one to two orders of magnitude higher. For the larger cluster in
the lower panel the individual steps are no longer resolved, resulting in a
peak of $\chi_P$ near $\omega=0.7\omega_{c}$.
  Since for the considered temperatures it
is unlikely to excite states with finite magnetic moment,
$\chi_P$ is small  at low $\omega$  .
It grows
with $\omega$ because these states come down.
It falls off at large $\omega$ when approaching
 the maximum magnetic moment of the electrons in the shell.
The curve  $T=0.1T_c$ shows still a double peak structure, which is residue
of the  discreteness of the electron levels.

\begin{figure}
\noindent
\mbox{\psfig{file=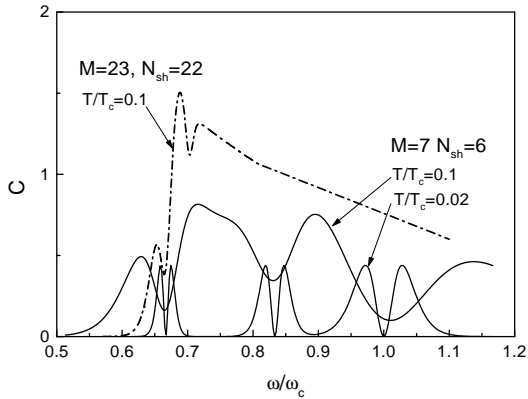,width=8cm}}
\caption{\label{f:com723} Heat capacity $C(T,\omega)$ v.s.  $\omega$. }
\end{figure}

The heat capacity is displayed  in fig. \ref{f:com723}.
Is has a double peak structure for the $M=7$ cluster at
 $T=0.02T_c$. In this case $C$ is very small because  the spacing between the
states of different $\nu$ is much larger than $T$.
Near a crossing the spacing becomes  small and $C$ goes up.  The dip
appears because at the crossing frequency the two states are degenerate.
 Then they do not contribute to $C$ because their
relative probability does not depend on $T$.
 For $T=0.1T_c$ the probability
to excite states with different seniority has increased and
$C$ takes substantial values between the crossings. The dips due to the
degeneracy  at the crossings remain.
For the  $M=23$ cluster $C$ shows only two wiggles, which are
 the residue of the discreteness of the electronic levels.

The deviation of real clusters from sphericity will attenuate the orbital
part of $\chi_P$ and round  the steps of $\Delta_c$
already  at $T=0$.  The back-bending phenomenon
observed in deformed rotating nuclei \cite{highk} is an example.
How strongly the orbital angular momentum is suppressed needs to be addressed
by a more sophisticated model than the present one.
In any case,  there will be steps  caused by the reorientation of
the electron spin, if the spin orbit coupling is small as in Al \cite{3}.
Most of the findings of the present paper are expected to hold qualitatively
for these spin flips.

In summary, at a temperature $T<0.1T_c$ an increasing external magnetic
field causes the magnetic moment of  small spherical
superconducting clusters ($R<5nm$) to grow in a step like manner.
Each step reduces the pair correlations until they are destroyed.
The steps manifest themselves as peaks in the magnetic susceptibility
and the heat capacity. The steps are washed out at $T>0.2T_c$.
For $T\sim T_c$, reduced but substantial pair correlations
persist to a higher field strength than for $T=0$.
This  phenomenon of the temperature-induced pairing in a strong
magnetic field is only found for the canonical ensemble.

Supported by the grant INTAS-93-151-EXT.

\end{document}